\documentclass{article}%
\usepackage{amsfonts}
\usepackage{amsmath}
\usepackage{amssymb}
\usepackage{graphicx}%
\providecommand{\U}[1]{\protect\rule{.1in}{.1in}}

\begin{document}

\title{Spectral - integral representation of the photon polarization operator in a
constant uniform magnetic field}
\author{V.M. Katkov\\Budker Institute of Nuclear Physics,\\Novosibirsk, 630090, Russia\\e-mail: katkov@inp.nsk.su}
\date{\ }
\maketitle

\begin{abstract}
The polarization operator in a constant and homogeneous magnetic field of
arbitrary strength is investigated on mass shell. The calculations are carried
out at all photon energies higher the pair creation threshold as well as lower
this threshold. The general formula for the effective mass of the photon with
given polarization has been obtained being useful for an analysis of the
problem under consideration as well as at a numerical work. Approximate
expressions for strong and weak fields $H,$ compared with the critical field
$H_{0}=4.41\cdot10^{13}%
\operatorname{G}%
,$ have been found. Depending on $H/H_{0}$ we consider the pure quantum region
of photon energy, where particles are created on lower Landau levels or
created not at all. Also the energy region of large level numbers is
considered where the quasiclassical approximation is valid.

\end{abstract}

1. The study of QED processes in a strong magnetic field close to and
exceeding the critical field strength $H_{0}=m^{2}/e=4,41\cdot10^{13}$ $%
\operatorname{G}%
$ (the system of units $\hbar=c=1$ is used ) is stimulated essentially by the
existence of very strong magnetic fields in nature. It is universally
recognized the magnetic field of neutron stars (pulsars) run up $H\sim
10^{11}\div10^{13}%
\operatorname{G}%
$ \cite{[1]}. These values of field strength gives the rotating magnetic
dipole model, in which the pulsar loses rotational energy through the magnetic
dipole radiation. The prediction of this model is in quite good agreement with
the observed radiation from pulsars in the radio frequency region. There are
around some thousand radio pulsars. Another class of neutron stars, now
referred to as magnetars \cite{[2]}, was discovered on examination of the
observed radiation at x-ray and $\gamma$-ray energies and may possess even
stronger surface magnetic fields $H\sim10^{14}\div10^{15}$ $%
\operatorname{G}%
$. The photon propagation in these fields and the dispersive properties of the
space region with magnetic is of very much interest. This propagation
accompanied by the photon conversion into a pair of charged particles when the
transverse photon momentum is larger than the process threshold value
$k_{\perp}>2m.$ When the field change is small on the characteristic length of
process formation (for example, when this length is smaller then the scale of
heterogeneity of the neutron star magnetic field), the consideration can be
realized in the constant field approximation. In 1971 Adler \cite{[3]} had
calculated the photon polarization operator in a magnetic field using the
proper-time technique developed by Schwinger \cite{[4]}. In the same year
Batalin and Shabad \cite{[5]} had calculated this operator in an
electromagnetic field using the Green function found by Schwinger \cite{[4]}.
In 1975 the contribution of charged-particles loop in an electromagnetic field
with $n$ external photon lines had been calculated in \cite{[6]}. For $n=2$
the explicit expressions for the contribution of scalar and spinor particles
to the polarization operator of photon were given in this work. For the
contribution of spinor particles obtained expressions coincide with the result
of \cite{[5]}, but another form is used.

The polarization operator in a constant magnetic field has been investigated
well enough in the energy region lower and near the pair creation threshold
(see, for example, the papers \cite{[7], [8], [9]} and the bibliography cited
there. In the present paper we consider in detail the polarization operator on
mass shell ( $k^{2}=0,$ the metric $ab=$ $a^{0}b^{0}-\mathbf{ab}$ is used ) at
arbitrary value of the photon energy and magnetic field strength. The
restriction of our consideration is only the applicability of the perturbation
theory over the electromagnetic interaction constant $\alpha$ \cite{[10]}$.$

2. Our analysis is based on the general expression for the contribution of
spinor particles to the polarization operator obtained in a diagonal form in
\cite{[6]} (see Eqs. (3.19), (3.33)). For the case of pure magnetic field we
have in a covariant form the following expression%

\begin{align}
\Pi^{\mu\nu}  &  =-\sum_{i=2,3}\kappa_{i}\beta_{i}^{\mu}\beta_{i}^{\nu
},\ \ \ \beta_{i}\beta_{j}=-\ \delta_{ij},\ \ \ \beta_{i}k=0;\label{1}\\
\beta_{2}^{\mu}  &  =(F^{\ast}k)^{\mu}/\sqrt{-(F^{\ast}k)^{2}},\ \ \ \beta
_{3}^{\mu}=(Fk)^{\mu}/\sqrt{-(F^{\ast}k)^{2}},\ \nonumber\\
\ FF^{\ast}  &  =0,\ \ \ F^{2}=F^{\mu\nu}F_{\mu\nu}=2(H^{2}-E^{2})>0,
\label{01}%
\end{align}
where $F^{\mu\nu}-$ the electromagnetic field tensor , $F^{\ast\mu\nu}-$ dual
tensor, $k^{\mu}$ $-$ the photon momentum, $(Fk)^{\mu}=F^{\mu\nu}k_{\nu},$%

\begin{equation}
\kappa_{i}=\frac{\alpha}{\pi}m^{2}r%
{\textstyle\int\limits_{-1}^{1}}
dv%
{\textstyle\int\limits_{0}^{\infty-\mathrm{i0}}}
f_{i}(v,x)\exp[\mathrm{i}\psi(v,x)]dx. \label{2}%
\end{equation}
Here%

\begin{align}
f_{2}(v,x)  &  =2\frac{\cos(vx)-\cos x}{\sin^{3}x}-\frac{\cos(vx)}{\sin
x}+v\frac{\cos x\sin(vx)}{\sin^{2}x},\nonumber\\
f_{3}(v,x)  &  =\frac{\cos(vx)}{\sin x}-v\frac{\cos x\sin(vx)}{\sin^{2}%
x}-(1-v^{2})\cot x,\nonumber\\
\psi(v,x)  &  =\frac{1}{\mu}\left\{  2r\frac{\cos x-\cos(vx)}{\sin
x}+[r(1-v^{2})-1]x\right\}  ;\label{3}\\
r  &  =-(F^{\ast}k)^{2}/2m^{2}F^{2},\ \ \ \mu^{2}=F^{2}/2H_{0}^{2}. \label{03}%
\end{align}
The real part of $\kappa_{i}$ determines the refractive index $n_{i}$ of the
photon with polarization $e_{i}=\beta_{i}$:%

\begin{equation}
\ \ \ \ n_{i}=1-\frac{\mathrm{\operatorname{Re}}\kappa_{i}}{2\omega^{2}}.
\label{4}%
\end{equation}
At $r>1$ the proper value of polarization operator $\kappa_{i}$ includes the
imaginary part which determines the probability per unit length of pair
production by photon with the polarization $\beta_{i}$:%

\begin{equation}
W_{i}=-\frac{1}{\omega}\mathrm{\operatorname{Im}}\kappa_{i} \label{5}%
\end{equation}
For $r<1$ the integration counter over $x$ in Eq. (\ref{2}) may be turn to the
lower semiaxis $(x\rightarrow-\mathrm{i}x),$ then the value $\kappa_{i}$
becomes real in the explicit form.

3. As well as in our work \cite{[11]} (see Appendix A) we present the
effective mass in the form of%

\begin{align}
\kappa_{i}  &  =\alpha m^{2}\frac{r}{\pi}T_{i};\ \ T_{i}=%
{\textstyle\int\limits_{-1}^{1}}
dv%
{\textstyle\int\limits_{0}^{\infty-\mathrm{i0}}}
f_{i}(v,x)\exp[\mathrm{i}\psi(v,x)]dx,\ \ \label{6}\\
T_{i}  &  =%
{\displaystyle\sum\limits_{n=0}^{\infty}}
\left(  1-\frac{\delta_{n0}}{2}\right)  T_{i}^{(n)};\ \ \ T_{i}^{(n)}=%
{\textstyle\int\limits_{-1}^{1}}
dv%
{\textstyle\int\limits_{0}^{\infty-\mathrm{i0}}}
F_{i}^{(n)}(v,x)\exp[\mathrm{i}a_{n}(v)x]dx, \label{7}%
\end{align}
where%

\begin{align}
F_{1}^{(n)}  &  =(-\mathrm{i})^{n}\exp(\mathrm{i}z\cot x)\left[
\frac{\mathrm{i}}{\sin x}(J_{n+1}(t)-J_{n-1}(t))-\frac{2vn}{z}\cot
xJ_{n}(t)\right]  ,\nonumber\\
F_{2}^{(n)}  &  =(-\mathrm{i})^{n}\exp(\mathrm{i}z\cot x)\frac{4}{z}\left(
b\cot x-\frac{\mathrm{i}}{\sin^{2}x}\right)  J_{n}(t)-F_{1}^{(n)},\nonumber\\
F_{3}^{(n)}  &  =F_{1}^{(n)}-2(-\mathrm{i})^{n}\exp(\mathrm{i}z\cot
x)(1-v^{2})\cot xJ_{n}(t);\label{8}\\
a_{n}(v)  &  =nv-b,\ \ b=\frac{1}{\mu}(1-r(1-v^{2})),\ \ z=\frac{2r}{\mu
},\ \ t=\frac{z}{\sin x}. \label{9}%
\end{align}
Let's note that at $x\rightarrow-\mathrm{i}\infty$ the asymptotic of the
Bessel function $J_{n}(t)$ is%

\begin{equation}
J_{n}(t)\simeq J_{n}(2\mathrm{i}ze^{-|x|})\simeq\frac{(\mathrm{i}z)^{n}}%
{n!}e^{-n|x|}, \label{10}%
\end{equation}
and under the condition $a_{n}(v)<n,$ the integration counter over $x$ in Eq.
(\ref{7}) can be unrolled to the lower semiaxis. Then $T_{i}^{(n)}$ becomes
real in the explicit form.

The functions $F_{i}^{(n)}(v,x)$ are periodical over $x.$ So one can present
$T_{i}^{(n)}$as%

\begin{align}
T_{i}^{(n)}  &  =%
{\textstyle\int\limits_{-1}^{1}}
dv%
{\textstyle\int\limits_{0}^{2\pi}}
F_{i}^{(n)}(v,x)\exp[\mathrm{i}a_{n}(v)x]dx%
{\displaystyle\sum\limits_{k=0}^{\infty}}
\exp[2\pi\mathrm{i}ka_{n}(v)]\nonumber\\
&  =%
{\textstyle\int\limits_{-1}^{1}}
\frac{dv}{1-\exp[2\pi\mathrm{i}a_{n}(v)]+\mathrm{i}0}%
{\textstyle\int\limits_{0}^{2\pi}}
F_{i}^{(n)}(v,x)\exp[\mathrm{i}a_{n}(v)x]dx. \label{11}%
\end{align}
Using the expression%

\begin{equation}
\frac{1}{1-\exp[2\pi\mathrm{i}a_{n}(v)]+\mathrm{i}0}=\frac{\mathcal{P}}%
{1-\exp[2\pi\mathrm{i}a_{n}(v)]}-\mathrm{i}\pi\delta(1-\exp[2\pi
\mathrm{i}a_{n}(v)]), \label{12}%
\end{equation}
taking into account the above notation%

\begin{align}
&  -\mathrm{i}\pi\delta(1-\exp[2\pi\mathrm{i}a_{n}(v)])\nonumber\\
&  =-\mathrm{i}\pi%
{\displaystyle\sum\limits_{m}}
\delta(1-\exp[2\pi\mathrm{i}(a_{n}(v)-m)])\rightarrow\frac{1}{2}%
{\displaystyle\sum\limits_{m\geq n}}
\delta(a_{n}(v)-m). \label{13}%
\end{align}
and allowing for $F_{i}^{(n)}(v,x+\pi)=(-1)^{n}F_{i}^{(n)}(v,x),$ we have%

\begin{align}
T_{i}^{(n)}  &  =(-1)^{n}\frac{\mathrm{i}}{2}\mathcal{P}%
{\textstyle\int\limits_{-1}^{1}}
\frac{dv}{\sin(\pi a_{n}(v))}%
{\displaystyle\int\limits_{-\pi}^{\pi}}
F_{i}^{(n)}(v,x)\exp[\mathrm{i}a_{n}(v)x]dx\nonumber\\
&  +%
{\displaystyle\sum\limits_{m\geq n}^{m=n_{\max}}}
{\displaystyle\sum\limits_{v_{1,2}}}
\frac{1+(-1)^{m+n}}{2|a_{n}^{\prime}(v)|}\vartheta(g(n,m,r))%
{\displaystyle\int\limits_{-\pi}^{\pi}}
F_{i}^{(n)}(v_{1,2},x)\exp[\mathrm{i}mx]dx, \label{14}%
\end{align}
where%
\begin{align}
g(n,m,r)  &  =r^{2}-(1+m\mu)r+n^{2}\mu^{2}/4,\ \nonumber\\
\ v_{1,2}  &  =\frac{n\mu}{2r}\pm\frac{1}{r}\sqrt{g},\ \ \ a_{n}^{\prime
}(v)\ =\frac{2}{\mu}\sqrt{g};\label{15}\\
n_{\max}  &  =[d(r)],\ \ \ d(r)=\frac{2(r-\sqrt{r})}{\mu}. \label{16}%
\end{align}
Here $[d]$ is the integer part of $d.$

Bringing out the distinction in the explicit form we present $T_{i}^{(n)}$ as%

\begin{align}
T_{i}^{(n)}  &  =T_{i}^{(nr)}+T_{i}^{(ns)};\label{17}\\
T_{i}^{(nr)}  &  =(-1)^{n}\frac{\mathrm{i}}{2}\mathcal{P}%
{\textstyle\int\limits_{-1}^{1}}
dv%
{\displaystyle\int\limits_{-\pi}^{\pi}}
[F_{i}^{(n)}(v,x)\frac{\exp[\mathrm{i}a_{n}(v)x]}{\sin(\pi a_{n}%
(v))}\nonumber\\
&  -%
{\displaystyle\sum\limits_{m\geq n}^{m=n_{\max}}}
{\displaystyle\sum\limits_{v_{1,2}}}
\frac{(-1)^{m}}{\pi}F_{i}^{(n)}(v_{1,2},x)\frac{\exp[\mathrm{i}mx]}%
{a_{n}(v)-m}]dx,\label{18}\\
T_{i}^{(ns)}  &  =%
{\displaystyle\sum\limits_{m\geq n}^{m=n_{\max}}}
{\displaystyle\sum\limits_{v_{1,2}}}
\frac{\mu\pi}{2\sqrt{g}}\left[  1-\frac{1}{\pi}\left(  \arctan\frac{2\sqrt
{-g}}{2r-\mu n}+\arctan\frac{2\sqrt{-g}}{2r+\mu n}\right)  \right] \nonumber\\
&  \times%
{\displaystyle\int\limits_{0}^{\pi}}
F_{i}^{(n)}(v_{1,2},x)\exp[\mathrm{i}mx]dx. \label{19}%
\end{align}
Here the regularized function $T_{i}^{(nr)}$ is singularity- free, and for
$n>n_{\max}\ $the integration counter in $T_{i}^{(n)}$ can be unrolled to the
lower semiaxis. . After that we present $T_{i}$ in the form%

\begin{align}
T_{i}  &  =%
{\displaystyle\sum\limits_{n>n_{\max}}^{\infty}}
T_{i}^{(n)}+%
{\displaystyle\sum\limits_{n=0}^{n_{\max}}}
T_{i}^{(n)}=\left(  T_{i}-%
{\displaystyle\sum\limits_{n=0}^{n_{\max}}}
T_{i}^{(n)}\right)  +%
{\displaystyle\sum\limits_{n=0}^{n_{\max}}}
T_{i}^{(n)}\label{20}\\
&  =%
{\textstyle\int\limits_{-1}^{1}}
dv%
{\textstyle\int\limits_{0}^{\infty}}
\{F_{i}(v,x)\exp[-\chi(v,x)]\label{21}\\
&  +\mathrm{i}%
{\displaystyle\sum\limits_{n=0}^{n_{\max}}}
F_{i}^{(n)}(v,-\mathrm{i}x)\exp[a_{n}(v)x]\}dx+%
{\displaystyle\sum\limits_{n=0}^{n_{\max}}}
T_{i}^{(n)}.\nonumber
\end{align}
Here the functions $F_{i}(v,x),\ \chi(v,x),\ \chi_{00}(v,x)$ have a form%

\begin{align}
F_{2}(v,x)  &  =\frac{1}{\sinh x}\left(  2\frac{\cosh x-\cosh(vx)}{\sinh^{2}%
x}-\cosh(vx)+v\sinh(vx)\coth x\right)  ,\label{22}\\
F_{3}(v,x)  &  =\frac{\cosh(vx)}{\sinh x}-v\frac{\cosh x\sinh(vx)}{\sinh^{2}%
x}-(1-v^{2})\coth x; \label{23}%
\end{align}%
\begin{align}
\chi(v,x)  &  =\frac{1}{\mu}\left[  2r\frac{\cosh x-\cosh(vx)}{\sinh
x}+(rv^{2}-r+1)x\right]  ,\label{24}\\
\chi_{00}(v,x)  &  =\frac{1}{\mu}\left[  2r+(rv^{2}-r+1)x\right]  . \label{25}%
\end{align}
and $a_{n}(v)$ is given by (\ref{9})

The integrals over $x$ in the expression for $T_{i}^{(ns)}$ (only there the
imaginary terms are contained) have been calculated in Appendix A \cite{[11]}.
Along with integers $m$ and $n$ we use also $l=(m+n)/2$ and $k=(m-n)/2$ which
are straight the level numbers%

\begin{equation}
r_{lk}=(\varepsilon(l)+\varepsilon(k))^{2}/4m^{2},\ \ \ \varepsilon
(l)=\sqrt{m^{2}+2eHl}=m\sqrt{1+2\mu l}. \label{026}%
\end{equation}
We have%

\begin{align}
\kappa_{i}^{s}  &  =\alpha m^{2}\frac{r}{\pi}%
{\displaystyle\sum\limits_{n=0}^{n_{\max}}}
\left(  1-\frac{\delta_{n0}}{2}\right)  T_{i}^{(ns)}=-\mathrm{i}\alpha
m^{2}\mu e^{-\zeta}%
{\displaystyle\sum\limits_{n,m}}
(2-\delta_{n0})\frac{\zeta^{n}k!}{\sqrt{g}l!}\nonumber\\
&  \times\left[  1-\frac{1}{\pi}\left(  \arctan\frac{2\sqrt{-g}}{2r-\mu
n}+\arctan\frac{2\sqrt{-g}}{2r+\mu n}\right)  \right]  D_{i};\ \ \ \label{27}%
\\
D_{2}  &  =\left(  \frac{m\mu}{2}-\frac{n^{2}\mu^{2}}{4r}\right)  F+2\mu
l\vartheta(k-1)\left[  2L_{k-1}^{n+1}(\zeta)L_{k}^{n-1}(\zeta)-L_{k}^{n}%
(\zeta)L_{k-1}^{n}(\zeta)\right]  ,\nonumber\\
D_{3}  &  =\left(  1+\frac{m\mu}{2}-\frac{n^{2}\mu^{2}}{4r}\right)  F+2\mu
l\vartheta(k-1)L_{k}^{n}(\zeta)L_{k-1}^{n}(\zeta),\nonumber\\
F  &  =\left[  L_{k}^{n}(\zeta)\right]  ^{2}+\vartheta(k-1)\frac{l}{k}\left[
L_{k-1}^{n}(\zeta)\right]  ^{2},\ \ \ \zeta=\frac{2r}{\mu}, \label{28}%
\end{align}
where $L_{k}^{n}(\zeta)$ is the generalized Laguerre polynomial.

At $\mu<<1,$ $(r-1)/\mu\lesssim1,$ $g/\mu\simeq|(r-1)/\mu-m|$ $<<1\ $the main
terms of sum in Eq. (\ref{53}) have a form:%

\begin{align}
\kappa_{3}^{s}  &  \simeq-\mathrm{i}\alpha m^{2}\mu e^{-\zeta}\zeta
^{m}g^{-1/2}%
{\displaystyle\sum\limits_{k+l=m}}
\frac{1}{k!l!}\label{29}\\
&  =-\mathrm{i}\alpha m^{2}\mu e^{-\zeta}\zeta^{m}g^{-1/2}\frac{2^{m}}%
{m!},\ \ \ \kappa_{2}^{s}\simeq\frac{1}{2}m\mu\kappa_{3}^{s}.\nonumber
\end{align}
Here we take into account that for $\zeta>>1$%

\begin{equation}
L_{k}^{n}(\zeta)\simeq\zeta^{k}/k!,\ \ \ D_{3}\simeq\left[  L_{k}^{n}%
(\zeta)\right]  ^{2},\ \ \ D_{2}\simeq m\mu D_{3}/2. \label{30}%
\end{equation}
Eq. (\ref{29}) coincides with Eq. (17) in \cite{[12]}. Note that for $g>0$ the
imaginary part of Eq. (\ref{27}) gives in addition the partial probability of
level population by created particles (see \cite{[11]}).

At $\mu\gtrsim1,$ $|r-1|\ <<1,$ $m=n=k=l=0,$ $D_{3}=1,$ $D_{2}=0$ and Eq.
(\ref{27}) coincides with Eqs. (15),(25) in \cite{[10]}. At $|$ $r-r_{10}%
|\ <<1$ the main term of sum is $m=n$ $=l=1,$ $k=0,$ $D_{2}=$ $\beta_{2},$
$D_{3}=$ $\beta_{3},$ $g=-h$ this equation coincides with Eqs. (33),(35) in
\cite{[10]}.

The work was supported by the Ministry of Education and Science of the Russian Federation.

\begin{quote}

\end{quote}


\begin{thebibliography}{99}                                                                                               %


\bibitem {[1]}M. Ruderman, in The Electromagnetic Spectrum of Nutron Stars,
NATO ANSI Proceedings (Springer, New York, 2004).

\bibitem {[2]}R.C. Duncan and C. Tompson, Astrophys. J. \textbf{392}, 19 (1992)

\bibitem {[3]}S.L. Adler, Ann. Phys. (N.Y.), \textbf{67}, 599 (1971).

\bibitem {[4]}J. Schwinger, Phys. Rev., \textbf{82, }664\textbf{
}(1951)\textbf{.}

\bibitem {[5]}I.A. Batalin and A.E.Shabad, Sov. Phis. JETP \textbf{33}, 483 (1971).

\bibitem {[6]}V.N. Baier, V.M. Katkov and V.M. Strakhovenko, Sov. Phis. JETP
\textbf{41}, 198 (1975).

\bibitem {[7]}V.N. Baier, A.I. Milstein and R.Zh. Shaisultanov, Zh. Eksp.
Teor. Fiz. \textbf{111}, 52 (1997).

\bibitem {[8]}A.C. Harding, M.G. Baring and P.L. Conthier, $\ $Astrophys. J.
\textbf{476}, 246 (1997).

\bibitem {[9]}A.E.Shabad, Zh. Eksp. Teor. Fiz. \textbf{125}, 210, (2004).

\bibitem {[10]}V.M. Katkov, arXiv:1403.3983 [hep-th] (2014).

\bibitem {[11]}V.N. Baier and V.M. Katkov, Phys. Rev., D \textbf{75}, 073009 (2007).

\bibitem {[12]}V.M. Katkov, arXiv:1311.6206 [hep-ph] (2013).
\end{thebibliography}
\end{document}